\documentclass[sigconf]{acmart}  

\usepackage{booktabs}
\usepackage{diagbox}
\usepackage{graphicx}
\usepackage{subcaption}
\usepackage{enumitem}
\usepackage{amssymb}
\usepackage{setspace}

\usepackage{tabularx}
\usepackage{dirtytalk}
\usepackage{arydshln} 

\usepackage{multirow}

\usepackage{enumitem}
\setlist[itemize,1]{leftmargin=0.4cm}

\begin{document}

\copyrightyear{2020} 
\acmYear{2020} 
\setcopyright{acmcopyright}\acmConference[CIKM '20]{Proceedings of the 29th ACM International Conference on Information and Knowledge Management}{October 19--23, 2020}{Virtual Event, Ireland}
\acmBooktitle{Proceedings of the 29th ACM International Conference on Information and Knowledge Management (CIKM '20), October 19--23, 2020, Virtual Event, Ireland}
\acmPrice{15.00}
\acmDOI{10.1145/3340531.3417433}
\acmISBN{978-1-4503-6859-9/20/10}

\fancyhead{}
\title{IAI MovieBot: A Conversational Movie Recommender System}

\author{Javeria Habib}
\affiliation{%
  \institution{University of Stavanger}
  \city{Stavanger}
  \country{Norway}
}
\email{javeriahabib09@gmail.com}

\author{Shuo Zhang}
\affiliation{%
  \institution{Bloomberg}
  \city{London}
  \country{United Kingdom}
}
\email{szhang611@bloomberg.net}

\author{Krisztian Balog}
\affiliation{%
  \institution{University of Stavanger}
  \city{Stavanger}
  \country{Norway}
}
\email{krisztian.balog@uis.no}

\begin{abstract}
Conversational recommender systems support users in accomplishing recommendation-related goals via multi-turn conversations. To better model dynamically changing user preferences and provide the community with a reusable development framework, we introduce IAI MovieBot, a conversational recommender system for movies. It features a task-specific dialogue flow, a multi-modal chat interface, and an effective way to deal with dynamically changing user preferences. The system is made available open source and is operated as a channel on Telegram.
\end{abstract}



\maketitle

\vspace*{-0.5\baselineskip}
\section{Introduction}

Conversational information access is a rapidly growing field that has been gaining attention over the past years~\citep{Anand:2020:CSD,Gao:2019:NAC}. A \emph{conversational recommender system} is a task-orientated system that supports its users in accomplishing recommendation-related goals through a multi-turn conversational interaction~\citep{Jannach:2020:SCR}. 

There exist a broad range of dialogue systems, from  ELIZA of the 60s~\cite{weizenbaum1966eliza} to modern-day social chatbots like Microsoft's XiaoIce~\cite{DBLP:XiaoIce}.
The task of conversational recommendation, however, is characterized by a set of unique needs. 
The agent needs to elicit the user's preferences, before 
it can make recommendations.  Further, these preferences may evolve or change during the course of interactions, as the user learns about the set of available items in the collection.
This desired behavior is currently not well supported in conversational information access systems.

While there are a number of open-domain dialogue frameworks to aid development, these are not suitable for the task at hand. 
OpenDial~\cite{DBLP:conf/acl/LisonK16} is designed to facilitate the development of agents for single-turn Q\&A-style dialogues. 
Plato~\cite{papangelis2019collaborative} and PyDial~\cite{Ultes:2017:PMS} attempt to model user preferences, but do not track the preference evolution over conversations. 
Further, most of the available domain-specific (movie) recommender systems are closed-source commercial products, such as the Facebook messenger bot And chill.\footnote{http://www.andchill.io/} Vote GOAT~\cite{Dalton:2018:VGC} is open-sourced, but its architecture comprises commercial components like Google Dialogflow, which require a paid plan.

In this work, we aim to address the above issues by developing an open-source conversational movie recommender system, which models users' preferences dynamically and supports multi-turn recommendations. 
Our system, termed IAI MovieBot, is based on an extensible architecture, comprising typical components, such as natural language understanding, dialogue manager, and natural language generation.
IAI MovieBot features several innovative elements:
\begin{itemize}
	\item A task-specific \emph{dialogue flow}, along with a set of associated \emph{intents}, to support the effective elicitation of user preferences and to provide suggestions from a large collection of items.
	\item A \emph{multi-Modal chat interface} to offer multiple modes of interactions and increase the ease of use of the system by allowing users to utilize on-screen buttons.		
	\item An effective way of dealing with dynamically changing \emph{user preferences}, which involves dialogue intents for revealing and modifying preferences, a machine-understandable structured representation of them, and an user interface and experience design where the agent makes the understood preferences transparent as well as scrutable, with the help of dedicated buttons.
\end{itemize}
As the name suggests, we operate in the movies domain. 
It should be noted, however, that the framework itself (i.e., the main components and dialogue intents) is domain independent and the specific system components can reasonably easily be adapted to other domains.


To summarize, the main contribution of this work is an open-source conversational movie recommender system featuring
(1) a task-specific dialogue flow, 
(2) a multi-modal chat interface, and
(3) an effective way to deal with dynamically changing user preferences.
The source code can be found at \url{https://github.com/iai-group/moviebot} and the system can be tried on the Telegram channel @IAI\_MovieBot.\footnote{\url{https://t.me/IAI_MovieBot}}
These resources are intended both for users acquiring movie recommendations and for researchers working on conversational recommender systems.

\section{Modeling and Architecture}
\label{sec:overview}

In this section, we present the overview of our framework for the task of conversational item recommendation. Specifically, we discuss how we model dialogue and user preferences (Sect.~\ref{sec:overview:modeling}) and describe the main architectural components (Sect.~\ref{sec:overview:architecture}). This framework is domain independent and is meant to be portable to other item recommendation scenarios beyond movies.

\begin{figure}[t]
\centering
\vspace*{-0.25\baselineskip}
\includegraphics[width=.56\linewidth]{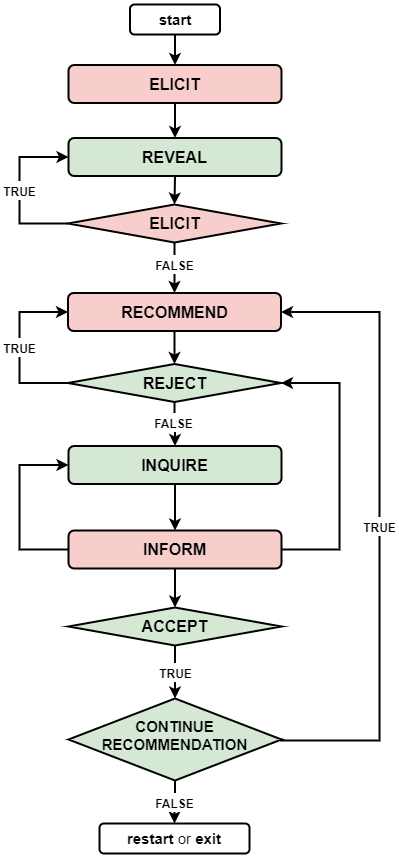}
\vspace*{-0.5\baselineskip}
\caption{Dialogue flow in IAI MovieBot. The colors \emph{red} and \emph{green} indicate agent's and users' intents respectively.}
\label{fig:IAI MovieBot_dialogue_flow}
\vspace*{-0.5\baselineskip}
\end{figure}

\subsection{Dialogue Modeling}
\label{sec:overview:modeling}

\begin{table}
\caption{Overview of intents.  Boldfaced intents can be further split into sub-intents (listed below the dotted line).}
\vspace*{-0.5\baselineskip}
\scriptsize
\label{tab:IntentsJarvis}
\begin{tabular}{p{.5cm}| p{1.5cm} p{5.5cm}}
\hline
& \textbf{Intent} & \multicolumn{1}{c}{\textbf{Description}}\\
\hline
& \multicolumn{2}{l}{\textbf{User Revealment~\cite{radlinski2017theoretical} / Query Formulation~\cite{trippas2018informing} / Query Reformulation~\cite{trippas2018informing}}} \\ \hline
\multirow{2}{0em}{User} & Reveal & The user wants to reveal a preference. \newline \emph{\say{Do you have any sports movies?}}\\
& Remove \mbox{preference} & The user wants to remove any previously stated preference. \newline \emph{\say{I don't want to see any sports movies anymore.}}\\ \hline
Agent & Elicit & Ask the user to describe their preferences. \newline \emph{\say{Which genres do you prefer?}}\\
\hline 
& \multicolumn{2}{l}{\textbf{System Revealment~\cite{radlinski2017theoretical} / Result Exploration~\cite{trippas2018informing}}} \\ \hline
\multirow{5}{0em}{User} & Inquire & Once the agent has recommended an item, the user can ask further details about it.\newline \emph{\say{Please tell me more about this movie.}}\\
\cline{2-3}
& \textbf{Accept/Reject} & The user can decide if they like the recommendation or not. \\ \cdashline{2-3}
& Accept & The user accepts (likes) the recommendation. This will determine the success of the system as being able to find a recommendation the user liked.\newline \emph{\say{I like this recommendation.}}\\
& Reject & The user either has already seen/consumed the recommended item or does not like it.\newline \emph{\say{I have already seen this one.}}\\
& Continue \mbox{recommendation} & If the user likes a recommendation, they can either restart, quit or continue the process to get a similar recommendation.\newline \emph{\say{I would like a similar recommendation.}}\\ \hline
\multirow{5}{0em}{Agent} & \textbf{Reveal} & Reveal the results or the number of matching results to the user.\\ \cdashline{2-3}
& Too many results & The number of items matching the user's preferences is larger than a maximum limit. This will be followed by an \emph{elicit} intent. \newline \emph{\say{There are almost 1100 action movies.}}\\
& Recommend & Based on the user's preferences, make a recommendation.\newline \emph{\say{I would like to recommend a fairy tale film, named Shrek.}}\\
& No Results & The database does not contain any items matching the user's preferences. \newline \emph{\say{Sorry. I couldn't find any romantic Korean movies.}}\\
& Inform & If the user \emph{inquires} about the recommended item, the agent provides the relevant information. \newline \emph{\say{The director of this movie is XYZ.}}\\
\hline
& \multicolumn{2}{l}{\textbf{Miscellaneous Intents}} \\ \hline 
\multirow{4}{0em}{User} & Hi & When the user initiates the conversation, they start with a formal \emph{hi/hello} or \emph{reveal} preferences.\\
& Acknowledge & Acknowledge the agent's question where required. \\
& Deny & Negate the agent's question where required.\\
& Bye & End the conversation by sending a bye message or an exit command.\\ \hline
\multirow{4}{0em}{Agent} & Welcome & Start the conversation by giving a short introduction. \\
& Acknowledge & Acknowledge the user's query where required. \\
& Cant Help & The agent does not understand the user's query or is not able to respond properly based on the current dialogue state.\\
& Bye & End the conversation.\\
\hline
\end{tabular}
\vspace*{-1\baselineskip}
\end{table}

We assume that recommendations happen via a multi-turn conversation with an agent, which is initiated and terminated by the user. 
The agent keeps eliciting user preferences until (a) the result set is sufficiently small or (b) it has reached the maximum amount of questions it is allowed to ask (to avoid fatiguing the user). Then, the agent makes recommendations for specific items and elicits feedback on them, until the user finds an item to their liking (or terminates the process).

User and agent utterances are modeled as \emph{dialogue acts}, which are machine-understandable representations of the natural language text. 
A dialogue act is mathematically represented as
\begin{equation*}
	\textit{intent}\big( 
	(\textit{slot}_1, \textit{op}_1, \textit{value}_1),
	\dots,
	(\textit{slot}_n, \textit{op}_n, \textit{value}_n)
\big ) ~,
\end{equation*}
%
where operators ($\textit{op} \in \{ =,\neq,<,>,\leq, \geq \}$) specify the relationship between for each slot and its corresponding value.
We create and define the set of possible intents based on \citep{radlinski2017theoretical} and \citep{trippas2018informing}; these are listed along with examples in Table~\ref{tab:IntentsJarvis}. 
Fig.~\ref{fig:IAI MovieBot_dialogue_flow} shows the dialogue flow in our system using these intents. 

The user's preferences are represented as an \emph{information need} (IN). The user can reveal their preferences at any stage in the conversation, which will trigger an update to the IN. 
Information needs are represented as slot-value pairs, and get their values assigned based on \emph{reveal} intents. For example, if a user wants a ``\textit{romance and comedy movie, starring Meryl Streep from the 90s},'' the IN will be modeled as \emph{[genres = romance, comedy; actors = Meryl Streep; release year $\geq$ 1990 \& < 2000]}.
Note that some slots can be multi-valued (this is defined by a domain-specific ontology).
Further, it may be that the system attempts to elicit preference for a slot that the user does not care about. Those responses are also registered, but they will not narrow the set of matching items. 

If the number of items matching the information need exceed a predefined threshold, the agent will attempt to elicit additional preferences (i.e., slot values for the IN). For example, if the user states a preference for \emph{action} movies, the agent will follow this up with the following request: ``\emph{There are almost 4700 action films. Please answer a few more questions to help me find a good movie...}''
As shown in Fig.~\ref{fig:IAI MovieBot_dialogue_flow}, once the agent has \emph{recommended} an item, the user has options to either \emph{reject} or \emph{accept} it, or \emph{inquire} further details about the item. 
These responses are recorded as \emph{dialogue context}. The dialogue context is a dictionary structure that stores all feedback associated with specific items, e.g., \emph{$\{$`Inception': [watched], `Amsterdamned': [dont\_like],`The Mountain II': [inquire, accept], ...$\}$}.
The user can receive more recommendations if they \emph{reject} an item and are offered similar recommendations if they \emph{accept}. They can also \emph{restart} or \emph{exit} a conversation. Restarting a conversation will erase the current IN as well as the history of recommended items.

\subsection{System Architecture}
\label{sec:overview:architecture}

\begin{figure}[t]
\centering
\vspace*{-0.25\baselineskip}
\includegraphics[width=\linewidth]{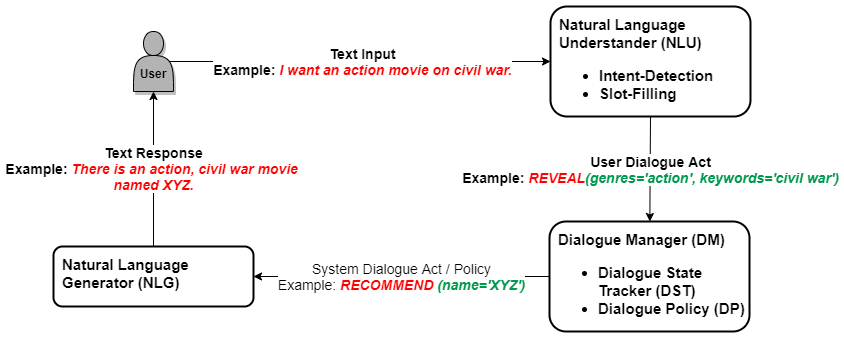}
\vspace*{-1.5\baselineskip}
\caption{IAI MovieBot system architecture.}
\label{fig:IAI MovieBot_ds_arch}
\vspace*{-0.75\baselineskip}
\end{figure}

The main architecture is shown in Fig.~\ref{fig:IAI MovieBot_ds_arch}, illustrating the core process for each dialogue turn. The \emph{natural language understander} (NLU) converts the natural language response from the user into a dialogue act. This process, comprising of \emph{intent detection} and \emph{slot filling}, is performed based on the current dialogue state. The \emph{dialogue state tracker} (DST) in the \emph{dialogue manager} (DM) updates the \emph{dialogue state} (DS) and \emph{dialogue context} (DC) based on the dialogue acts by both the agent and the user.  
The dialogue state includes the recent dialogue acts for both the user and the agent, the information need, the matching results with respect to the IN, the current recommendation by the agent, and the agent's state that defines its next step. The dialogue context keeps track of the items recommended to the user with their feedback (where possible values include ``accepted,'' ``rejected/don't like,'' and ``inquire'').
The dialogue policy (DP) generates a dialogue act by the agent based on the current dialogue state. It defines the flow of the conversation, i.e., what steps an agent must take at every stage. For example, the intent \emph{elicit} is generated if the IN contains less than two user preferences. If the IN has at least one value, the intent \emph{too many results} is generated. The parameters of the dialogue act represent what the agent must elicit, recommend, or inform. At any stage, a \emph{reveal} or \emph{remove preference} intent by the user will lead to \emph{recommend} or \emph{elicit}. For the intent \emph{recommend}, the dialogue policy triggers the generation of an item recommendation from the collection. The output of the DP is converted to a natural language response by the \emph{natural language generator} (NLG). The NLG also summarizes the IN back to the user, to help them keep track of their stated preferences. Further, the NLG helps the user to  explore the item space by providing options. 

\section{Implementation}

IAI MovieBot is implemented in Python as a client-server application.  This section highlight some of the domain-specific aspects; for specific details, we refer to the GitHub repository.

\subsection{Item Collection}
We base our item collection on the publicly available MovieLens 25M Dataset.\footnote{https://grouplens.org/datasets/movielens/25m/} 
This dataset contains 62k movies with 25M ratings and 1M tags assigned by 163k users. We use IMDbPY\footnote{https://imdbpy.github.io/} to retrieve movie details from IMDb, including \emph{genres}, \emph{movie keywords}, list of \emph{actors}, \emph{directors}, movie \emph{duration}, \emph{plot (summary)}, \emph{release year}, IMDb \emph{rating} with \emph{number of votes} and \emph{links} to the movie page and its cover image on IMDb.  After filtering  movies that are missing essential attributes, we end up with around 40k movies, which we store in a MySQL database.

\subsection{Natural Language Understanding}
\vspace*{-0.1\baselineskip}

\subsubsection{Intent detection.} 
We use a pattern-based method for intent detection, where a matched intent with a high probability is considered as the output. For example, \emph{inquire} happens mostly when the agent has recommended a movie. 
To check if the user is inquiring or revealing a \emph{director name}, the response should have one of the patters \emph{``directed,''} \emph{``director,''} or \emph{``directors.''} 

\subsubsection{Slot filling.}
A list of possible values is created for each slot using the underlying database, then matching is performed on lemmatized word forms. However, for recognizing \emph{genres}, synonyms are also considered. 
For \emph{movie keywords} and \emph{title}, we operate on n-grams ($n=8...1$). For each n-gram, the substring is lemmatized, and it is matched to the lemmatized slot-values. 
The detection of \emph{release years} employs a set of specific patters to be able to deal with values like \emph{``90s,''} \emph{``1950s,''} \emph{``1995,''} or \emph{``20th century.''} 
Annotations are further filtered to ensure that the same text span is not annotated for multiple slots. For example, given the user utterance \emph{``I want movies on the civil war,''} the initial slot filling will yield \emph{genres=war} and \emph{keywords=civil war}. Therefore, the filtered results will exclude the \emph{genres} slot as the word \emph{war} is also present in \emph{keywords}.
Further, slot filling also detects preference statements. This is done by pattern matching for the set of words preceding the detected value. For example, \emph{``I want action movies but not directed by Brad Pitt,''} the resulting values are \emph{genres=action} and \emph{director $\neq$ Brad Pitt}. For detecting a person name, which can be both a director or an actor in the collection, we use pattern matching to check if a preference is mentioned, e.g., \emph{directed by} or \emph{starring}. It is also considered that the user may prefer a movie \emph{before} or \emph{after} a specified time period.

\subsection{Natural Language Generation}
\vspace*{-0.1\baselineskip}

To avoid robotic responses, multiple templates are designed for each intent and its parameters to select a response randomly. For example, for a dialogue act \emph{elicit(keywords)}, the response will be generated randomly from (i) \emph{``Can you give me a few keywords?''} and (ii) \emph{``What are you looking for in a movie? Some keywords would be good.''} Moreover, for dialogue act \emph{inform(director = Jennifer Lee)}, the response can be either (i) \emph{``The director of this movie is Jennifer Lee.''} or (ii) \emph{``Its directed by Jennifer Lee.''}

\subsection{Dialogue Manager}
\vspace*{-0.1\baselineskip}

The dialogue state tracker updates the dialogue state and dialogue context in three stages: (i) the DM receives the user's dialogue acts from the NLU; (ii) candidate recommendations that match the user's preferences as represented in the IN are generated from the item collection; (iii) the DP generates the agent's dialogue acts as output.

The dialogue policy takes the conversation history into account and filters out movies that have already been recommended (i.e., those that are stored in the dialogue context). 

\begin{figure}[t]
  \centering
  \begin{tabular}[b]{c c}
    \includegraphics[width=.49\linewidth]{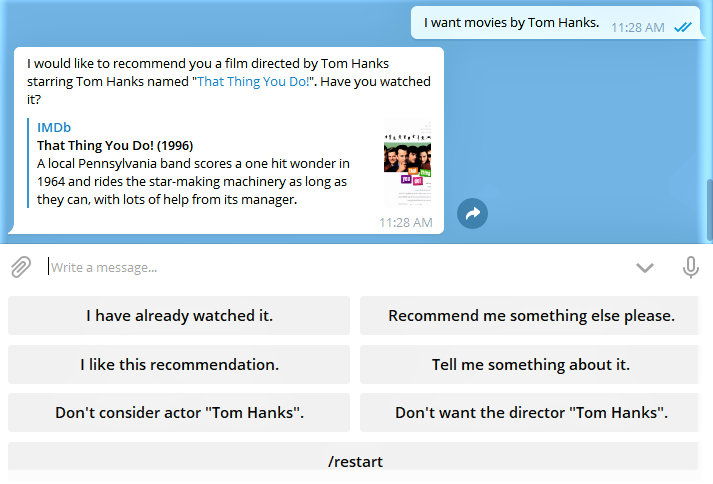} &
    \includegraphics[width=.49\linewidth]{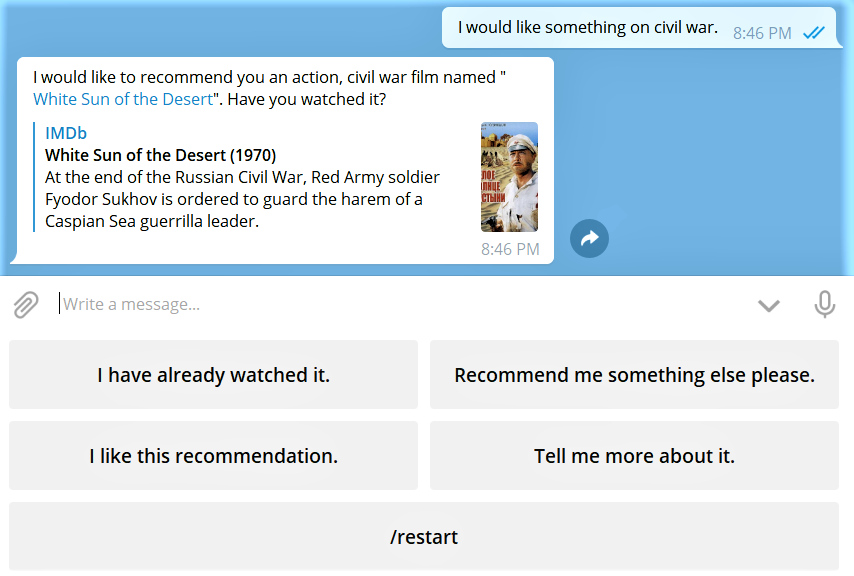} \\
    \small (a) & (b) \\
    \includegraphics[width=.49\linewidth]{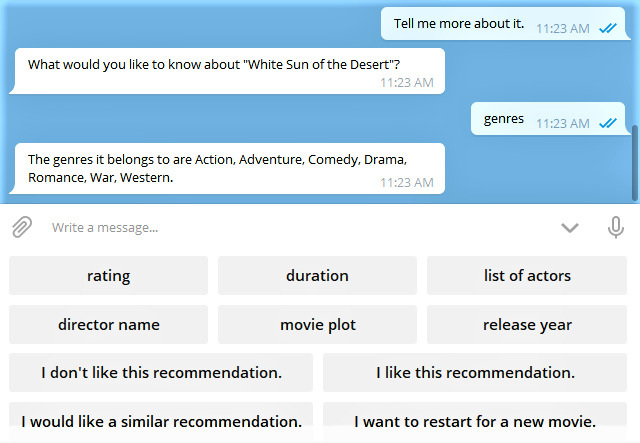} &
    \includegraphics[width=.49\linewidth]{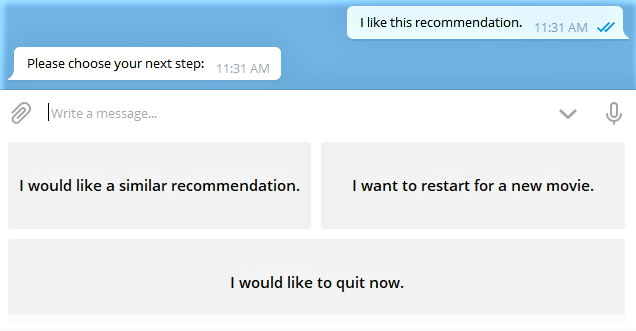} \\
    \small (c) & (d)
  \end{tabular}
\vspace*{-1\baselineskip}
\caption{Telegram keyboard buttons (a) the information need has the same value (Tom Hanks) for two slots (actors and directors), (b) a recommendation is made, and the options are accepted/reject and inquire, (c) the user has inquired about the movie genres, and remaining attributes are presented as buttons, (d) continue recommendation. }
\label{fig:buttons}
\vspace*{-1\baselineskip}
\end{figure}

\subsection{Multi-modal Chat Interface}

The chat interface is implemented on Telegram.\footnote{\url{https://telegram.org/}}
For the Telegram integration, the python-telegram-bot API is used.\footnote{\url{https://github.com/python-telegram-bot/python-telegram-bot}} The user's options, generated by the NLG, are shown as keyboard buttons in the Telegram app. The text of each button corresponds to a possible response and is linked to a specific dialogue act. The following intents are shown as buttons:

\begin{itemize}
    \item \emph{Remove Preference}: If two slots in the IN have the same value, it is assumed that the user intends to assign the value to one of these slots. Therefore, the buttons are shown for the user to ease the removal one of their preferences, if needed (Fig \ref{fig:buttons} (a)). This ambiguity typically occurs if a person name is detected in the utterance.
    \item \emph{Accept/Reject}: The options to \emph{accept} and \emph{reject} a movie are presented when the agent has made a recommendation (Fig \ref{fig:buttons} (a, b)) or is informing the user about it (Fig \ref{fig:buttons} (c)). The button to \emph{continue recommendation} is presented while the agent is informing the user about the movie (Fig \ref{fig:buttons} (c)) or the user likes the recommendation and may want to find similar items (Fig \ref{fig:buttons} (d)).
    \item \emph{Inquire}: When the agent recommends a movie, the user may want to inquire further about it. The buttons for inquiring about a movie are split into two steps: 
		(1) One button stating that the user wants to know more about the movie (Fig \ref{fig:buttons} (a, b)).
        (2) Buttons representing the movie attributes that can be inquired about. Each button gets be removed once the user has asked about that attribute (Fig \ref{fig:buttons} (c)).
\end{itemize}

\noindent
Moreover, commands to get \textit{help}, and \textit{start}, \textit{restart}, and \textit{exit} the conversation are added to the interface.

\subsection{Feedback Collection}

To help improve the system, a feedback form is created.\footnote{\url{https://forms.gle/q4m5fwaWMZigVsaP7}}
Users fill out this form anonymously, by providing only basic demographic information (age, gender, and education). Survey respondants are asked to rate the system based on quality attributes: \emph{effectiveness}, \emph{efficiency}, and \emph{satisfaction}~\cite{DBLP:journals/corr/RadziwillB17}. 
They are further invited to provide free text feedback on what they liked and disliked most about the system. 
The feedback we solicited during various stages of development helped us to shape and improve the system's functionality.

\section{Conclusion}

We have presented IAI MovieBot, an open-source conversational recommender system for movies. 
The system follows a task-specific dialogue flow, in which user preferences are elicited until the set of matching items is sufficiently small to make effective recommendations. The user experience has been designed to cater for dynamically changing preferences.
IAI MovieBot offers a multi-model chat interface and is made available as a channel on Telegram. 

According to the feedback we have received from users, IAI MovieBot has proved to be successful in understanding their preferences, helping them to grasp their options during various stages of the conversation, and ultimately recommending a good movie.
Nevertheless, the current system is seen as a starting point and we envisage expanding it further. The NLU and NLG components employ simple rule/template-based solutions, which we aim to replace with more advanced (neural) methods in the future. 
We also wish to extend the system to other domains.




\bibliographystyle{ACM-Reference-Format}
\bibliography{00paper}

\end{document}